# Deep learning-based algorithm for assessment of knee osteoarthritis severity in radiographs matches performance of radiologists

Albert Swiecicki, Nianyi Li, Jonathan O'Donnell, Nicholas Said, Jichen Yang, Richard C. Mather, William A. Jiranek, Maciej A. Mazurowski

## Abstract

A fully-automated deep learning algorithm matched performance of radiologists in assessment of knee osteoarthritis severity in radiographs using the Kellgren-Lawrence grading system.

## Purpose

To develop an automated deep learning-based algorithm that jointly uses Posterior-Anterior (PA) and Lateral (LAT) views of knee radiographs to assess knee osteoarthritis severity according to the Kellgren-Lawrence grading system.

## Materials and Methods

We used a dataset of 9739 exams from 2802 patients from Multicenter Osteoarthritis Study (MOST). The dataset was divided into a training set of 2040 patients, a validation set of 259 patients and a test set of 503 patients. A novel deep learning-based method was utilized for assessment of knee OA in two steps: (1) localization of knee joints in the images, (2) classification according to the KL grading system. Our method used both PA and LAT views as the input to the model. The scores generated by the algorithm were compared to the grades provided in the MOST dataset for the entire test set as well as grades provided by 5 radiologists at our institution for a subset of the test set.

## Results

The model obtained a multi-class accuracy of 71.90% on the entire test set when compared to the ratings provided in the MOST dataset. The quadratic weighted Kappa coefficient for this set was 0.9066. The average quadratic weighted Kappa between all pairs of radiologists from our institution who took a part of study was 0.748. The average quadratic-weighted Kappa between the algorithm and the radiologists at our institution was 0.769.

## Conclusion

The proposed model performed demonstrated equivalency of KL classification to MSK radiologists, but clearly superior reproducibility. Our model also agreed with radiologists at our institution to the same extent as the radiologists with each other. The algorithm could be used to provide reproducible assessment of knee osteoarthritis severity.

# Introduction

Tibiofemoral osteoarthritis is defined as loss of articular cartilage from the medial and/or lateral femoral condyles and patella femoral articular surfaces from mechanical overload due to angular or post traumatic changes, combined with the development of osteophytes, subchondral sclerosis, and periarticular cyst formation.

In the United States, Osteoarthritis accounts for $80 billion in annual direct costs (Agency for Healthcare Research and Quality, 2017) with more than 30 million Americans affected (CDC, 2020). Total Knee Arthroplasty (TKA) is an effective surgical intervention for knee osteoarthritis (KOA), but the procedure is costly, and its utilization is quickly increasing. In 2014, more than 700,000 inpatient knee replacements were performed at an average charge of $57,500 (Healthcare Cost and Utilization Project, n.d.). Conservative projections forecast a 143% increase in knee replacement utilization by 2050 (Inacio et al., 2017), attributable to an aging population with longer life expectancy (Losina et al., 2012).

In the setting of value-based payment reform (Porter, 2010), appropriate use of expensive procedures, including TKA, is paramount.. One study (Riddle et al., 2015) notably found that approximately one-third of TKA in a US sample were "inappropriate" using published appropriateness criteria . Several "appropriateness criteria have been proposed, all of which include radiographic evaluation of KOA severity (Riddle et al., 2014) (American Academy of Orthopaedic Surgeons, n.d.).

The Kellgren-Lawrence (KL) assessment is a commonly used and accepted KOA severity scale from 0 to 4 that incorporates a radiologist's measurement of joint space narrowing (JSN), osteophyte description, and bony changes such as sclerosis and deformity in knee radiographies, including Posterior-Anterior (PA) and Lateral (LAT) views (KELLGREN & LAWRENCE, 1957). Through current assessment processes, human error introduces inter-reader variability (Riddle et al., 2013) into scoring KOA severity. As a central component of appropriate use criteria, KOA severity, such as through KL assessment, will need to be reproducibly obtained, uniformly processed, and easily accessible to all healthcare stakeholders.

An algorithmic assessment of could be of great utility in the process of osteoarthritis case. In primary care practices, it could help the providers with a proper referral of a patient with knee pain. In radiology, it would add the KL assessment where was not made (a frequent occurrence). The assessment provided by a computer algorithm would address the issue of intra- and inter-reader variability since one algorithm could assess large numbers of studies across different institutions and would always return the same value for the same image. Such assessment could then be relied upon in an orthopedic surgery practice to make a decision regarding surgical treatment.

In recent years deep learning has dominated the OA severity estimation task. In two publications a) (Tiulpin et al., 2018) and b) (Antony et al., 2017) OA assessment task was divided into two stages of detection and classification where detection was performed using Faster R-CNN (Ren et al., 2017) and classification was done by deep convolutional networks. The third publication (Thomas et al., 2020) does not mention detection procedure and use deep convolutional networks for classification purposes. These publications use MOST (Segal et al., 2013) and Osteoarthritis Initiative (OAI) (Eckstein et al., 2012) datasets. The prior studies proposed separate channel for medial and lateral knee compartments using fusion of Siamese networks models, and transfer learning (Pan & Yang, 2010) approach fine-tuning different versions of ResNet architectures (He et al., 2016) pre-trained on ImageNet dataset (Jia Deng et al., 2009) as well as training with dual objective function combining classification and regression tasks at the same time, higher KL grade corresponds to more severe osteoarthritis extent. However, the prior studies did not incorporate the lateral radiographic view in the assessment of KL grade which is as a standard component of this assessment in the MOST study. Furthermore, prior studies did not offer a comparison to multiple radiologists in order to evaluate performance of the algorithm in a real clinical setting.

In this study we present state-of-the-art results in automatic KL assessment. Our model incorporates lateral radiographs (LAT) into the view into the KL grade prediction. Additionally, we perform a reader study with five radiologists in order to evaluate our model in a realistic clinical setting.

# Methods

*Dataset*

In this study, we used the dataset from The Multicenter Osteoarthritis Study (MOST, https://most.ucsf.edu/). The study collected clinical assessment and radiological data at multiple time points. We used weight-bearing PA and LAT view knee radiographs from MOST. The original dataset consists of data for 10052 exams for 3026 patients. MOST has completed baseline, 15-, 30-, 60-, 72- and 84- month follow-up visits. Kellgren-Lawrence grade was assigned to the knees in this dataset in the following way:

grade 0: no radiographic features of OA are present

grade 1: doubtful joint space narrowing (JSN) and possible osteophytic lipping

grade 2: definite osteophytes and possible JSN

grade 3: multiple osteophytes, definite JSN, sclerosis, possible bony deformity

grade 4: large osteophytes, marked JSN, severe sclerosis and definite bony deformity

The dataset also contained joint space narrowing (JSN) score and osteophyte grade assessments.

Given this original dataset, we performed some exclusions to arrive at a final set used for the analysis. If a knee in a given visit was specially marked in the dataset (for example, data is missing or poor-quality), such knee was not considered in our analysis for that visit. For example, if a patient had TKA performed on right knee after V2, we included right and left knee images from V0, V1, V2 and included only left knee's images from all visits after V2. We also did not consider a knee in a given visit if for that visit the knee was missing JSN or osteophyte grades, since our further study will include JSN and osteophytes estimation and this process only excludes less than 10% of all samples.

Finally, if for a given visit paired PA and LAT views were missing, we also did not consider such knee for that visit. In the case of multiple images for one view available from the same patient visit, we selected one randomly. The final dataset used in our study contained 9739 exams for 2802 patients. This resulted in data for 18053 knee appearances with 9739 PA, 9239 left LAT and 9264 right LAT images.

We randomly split the remaining data by assigning patients into non-overlapping training, validation and test subsets. Dataset across sets before detection is shown in table 1. KL grade distribution over the set is presented in table 2.

| Classification set | Training | Validation | Test |
|---|---|---|---|
| Number of patients | 2040 | 259 | 503 |
| Number of exams | 7062 | 914 | 1763 |
| Number of knees | 13404 | 1740 | 3359 |

Table 1. Dataset split

| Set\KL grade | 0 | 1 | 2 | 3 | 4 |
|---|---|---|---|---|---|
| Train | 5600 | 1951 | 2228 | 2475 | 1150 |
| Valid | 678 | 263 | 329 | 316 | 154 |
| Test | 1283 | 526 | 588 | 697 | 265 |

Table 2. KL grades distribution between train, validation and test sets

Since grade 1.9 is not a part of the standard KL grading system in experiments we treated grades 1.9 as grade 2 (according to the MOST dataset description and reading protocol from July 2016).

*Knee joint annotations*

In order to obtain data for training of a joint detection model, we manually pointed out the center of the knee joint, which can represent the center of the bounding box, for randomly selected 600 knees from the training set (300 left and 300 right). For each knee, the joint was pointed out in the PA and the LAT view resulting in the total of 1200 annotations. Then, the data was randomly divided into 210 knees for training of the detection model, 45 knees for the validation of the model, and 45 knees for the testing of the detection model.

*Kellgren-Lawrence annotations from our institution*

In addition to the KL score annotations available in the MOST dataset, we collected labels for 204 randomly selected knee images from the classification test set from 5 radiologists at our institution. It resulted in 1020 unique labels. Radiologists assigned KL grades base on PA and LAT views.

The images were graded by 4 board-certified Radiology Fellows/Clinical Instructors and one attending Radiologist with 5 years post fellowship experience in the Musculoskeletal Division of the Duke Department of Radiology   in the Musculoskeletal Division Duke Department of Radiology.

*Image preprocessing*

Images read from DICOM (Mildenberger et al., 2002) files were resized according to the "PixelSpacing" tag using bicubic interpolation algorithm into uninformed resolution corresponding to 0.2 mm per pixel since it is the most common value across DICOM images in the dataset. Afterwards, we converted images into 8-bit grayscale deep from 16-bits. Finally, each image was normalized by dividing by maximum pixel intensity and subtracting of pixel intensity standard deviation.

*The algorithm for knee osteoarthritis severity estimation*

Our approach consists of two steps: (1) detecting knee joints and (2) classifying knee joints in terms of OA. Together those two steps constitute a fully automated pipeline allowing to estimate KL grade. Figure. 1 Shows entire pipeline.

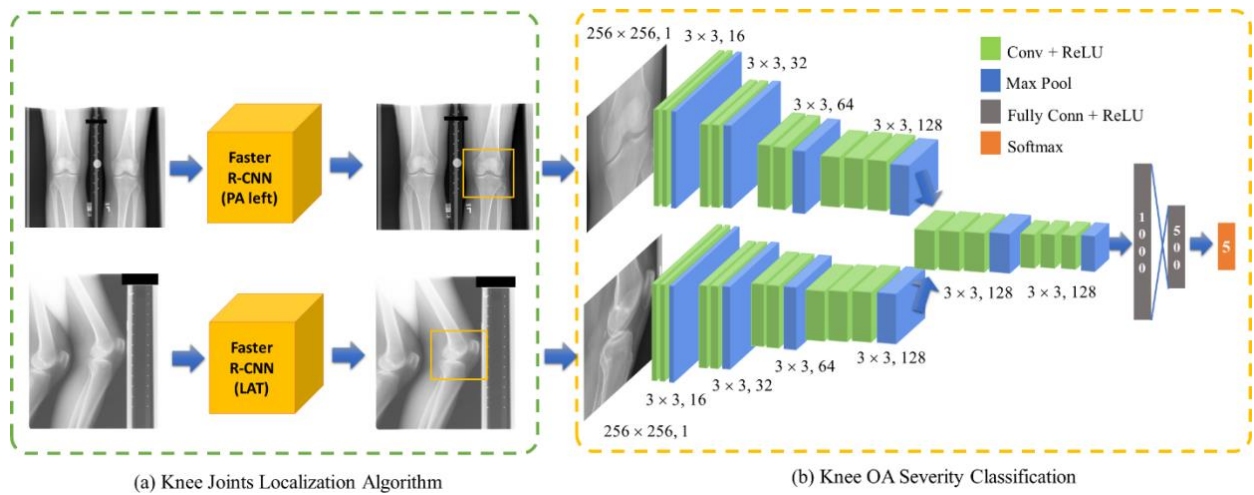

Figure 1. The classification pipeline

*Step 1: knee joint localization algorithm*

Inspired by the success of the Region Proposal Network (RPN) (Ren et al., 2017) that shares full-image convolutional features with the detection network, we utilized the RPN to automatically detect the region of interest (ROI) within X-ray images. An RPN is a fully convolutional network that simultaneously predicts object bounds and objectless scores at each position, which are used by Faster R-CNN for detection. The RPN is trained end-to-end to generate high-quality region proposals. As a Faster R-CNN backbone, we used VGG16 architecture.

As training labels, we provided images bounding boxes for previously labeled images around the center of the knee join with height and width of 1000 pixels (corresponding to 200 mm). As the loss function that measures detection quality, we utilized the Jaccard index (Crum et al., 2006) also known as Intersection over Union (IoU). We trained the detection models separately right and left side of the LAT and the PA views. When running of the model resulted in multiple detection for each knee, we used the one with the highest model certainty. As a result of detection, we stored the centers of predicted bounding boxes.

In order to evaluate the detection model, we calculate the proportion of knees that reached IoU $\geq$ 0.75.

Using transfer learning (Pan & Yang, 2010) we utilized weights pretrained on ImageNet VGG16 and we fine-tuned. Models were trained with batch size of 4 images, Adam optimizer with learning alpha equal to 0.0001 and early stopping after 20 epochs without increased Jaccard index value on validation set.

Detection process is shown in Figure. 2.

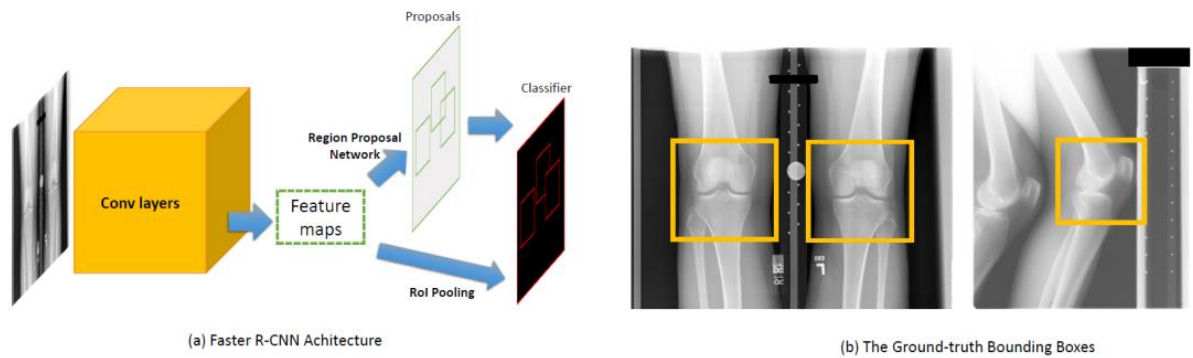

Figure. 2 Faster R-CNN for knee joints localization. (a) The Faster R-CNN architecture. (b) A knee OA X-ray image with the region of interest. From left to right: PA radiograph with bounding boxes; LAT radiograph of a left knee with bounding box.

*Step 2: Knee OA severity classification using Multi-input CNN*

Unlike prior approaches to this task, our approach is based on a multi-input CNN architecture, which takes the PA and LAT view into account simultaneously. As shown in Fig. 1, the entire network consists of two branches in the early layers, one corresponding to each of the input images. Each branch consists of convolution, max-pooling, batch-normalization and rectified linear unit (ReLU) layers. We concatenate channels outputs and feed the combined feature maps to the top convolutional layers. On top of the convolutional layers, we add two fully connected layers to predict the KL scores representing the OA severity. The output of the model is a vector of KL grade confidence scores. As a final result, we take prediction with the highest confidence score.

The input to the networks is a PA and LAT 700 x 700 pixels (140 x 140 mm) patches cropped from radiographs around the knee joint center. All knee joint centers were estimated using the Faster R-CNN model described in section 3.2. Cropped patches were resized to 256x256 pixel patches.

We used PyTorch as the software framework and 2× Nvidia GTX1080 cards with 8GB memory for each experiment. We trained the model using Adam optimizer with learning rate 0.00001 and set early stopping patience to 20.

To avoid issues with bias in early epochs of adaptive moment estimation (ADAM) optimizer (Kingma & Ba, 2015) in early epochs with we decided to train a model for the first 10 epochs without any data augmentation. In the following epochs, we used data augmentation on training data consist of horizontal flips, rotations, translations, scaling and shearing.

In order to overcome an issue with an unbalanced dataset (Buda et al., 2018), we decided to use a method similar to oversampling, which rely on sampling cases for each batch from the train set with the same probability for each of the all classes.

The validation set was adjusted using the same method to prevent from favoring classes with more samples. It was created once and remained constant for all epochs of the described experiments. Since during experiments, we observed that random initialization has a significant impact on obtained validation score (probably due to finding different local minima) we decided to train 10 models in the same way and use the one which obtained the best validation score. After selecting the model with the highest validation score, we tested our model on the test set.

Additionally, we developed single view architectures for PA and LAT views separately. The architectures in both cases are the same. Architecture details are presented in appendix A. The training process of the single-view model is identical to the multi-input model.

*Assessment of the proposed models in a multi-reader setting*

To provide a final evaluation of the developed models we assessed the agreement of our algorithm with each of the 5 radiologists at our institution and compared to the agreement between radiologists themselves.

In order to assess agreement between pairs of readers as well as between radiologists and our deep learning algorithm, we used a quadratic-weighted Kappa coefficient. It reflects the agreement between two raters and weighs the different misclassification errors differently. The same metric was used in a previous study on KL scores (Tiulpin et al., 2018)In appendix B we also provide a non-weighted Kappa coefficient and lineal weighted Kappa coefficient.

## Results

*Knee Joint detection*

We are evaluating our detection algorithm using the IoU metric, which is the intersection area of the annotated bounding box and the predicted bounding box divided by the union area of the two boxes. Our faster R-CNN models found the knee joints in 99.44% of the knees at the IoU>=0.75. The mean IoU was equal to 92.24 with a standard deviation (STD) of 4.15. Only one knee was detected with IoU<0.75 (IoU=0.73). Detailed metrics for respective models are shown in table 3.

|  | PA right | PA left | LAT right | LAT left |
|---|---|---|---|---|
| IoU mean +/- stdev | 93.73 +/- 3.70 | 92.23 +/- 3.73 | 92.30 +/- 3.73 | 90.71 +/- 4.80 |

Table 3. Detection results on detection test set

*Knee osteoarthritis diagnosis*

The multi-input model yielded 71.90% accuracy on the entire test set. The confusion matrix and the normalized confusion matrix are shown in Figure 3. The detailed comparison of classification results is showed in table 4. The quadratic weighted Kappa coefficient between our multi-input model predictions and annotations from MOST community clinical experts on the test set is equal to 0.907.

Single view models yielded accuracies of 70.85% and 61.18% for PA and LAT views respectively. The confusion matrices for single view models are shown in appendix B.

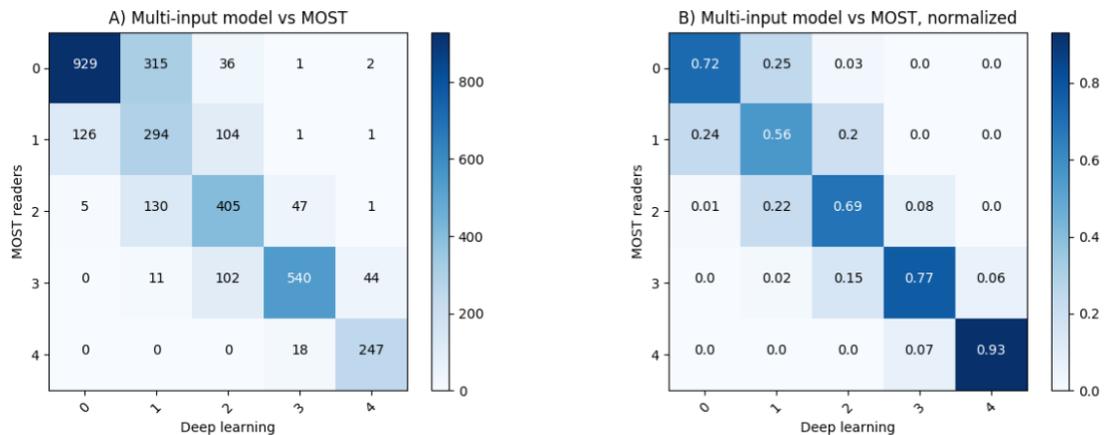

Figure 3. The Multi-input model confusion matrix A) Unnormalized B) Normalized.

| Study | Training data | Validation data | Test data | Detection | Methodology | Accuracy |
|---|---|---|---|---|---|---|
| Antony, et al., 2017 | MOST + OAI | MOST + OAI | MOST + OAI | Manual | Classification only | 61.9% |
| Antony, et al., 2017 | MOST + OAI | MOST + OAI | MOST + OAI | Faster R-CNN | Classification and Regression | 63.6% |
| Tiulpin, et al., 2018 | MOST | OAI | OAI | Faster R-CNN with manual corrections | Separate channels for lateral and medial compartments and model fusion | 66.71% |
| Tiulpin, et al., 2018 | MOST | OAI | OAI | Faster R-CNN with manual corrections | Fine-tuned pretrained ResNet-34 (Transfer learning) | 67.49% |
| Thomas et al., 2020 | OAI | OAI | OAI | Unknown | Fine-tuned pretrained ResNet-169 (Transfer learning) | 70.66% |
| Our | MOST | MOST | MOST | Faster R-CNN | LAT view | 61.18% |
| Our | MOST | MOST | MOST | Faster R-CNN | PA view | 70.85% |
| Our | MOST | MOST | MOST | Faster R-CNN | PA and LAT channels | **71.90%** |

Table 4. Comparison of classification results

*Classification performance in comparison to 5 readers at our institution*

The table of quadratic-weighted Kappa's is shown in Fig. 4. The average quadratic weighted Kappa coefficient between all pairs of radiologists at our institution was 0.748. The average quadratic-weighted Kappa between Duke Hospital readers and MOST readers was 0.759. The average quadratic-weighted Kappa between our multi-input model and the 5 readers at our institution was 0.769. Table representing regular Kappa is included in appendix C. Confusion tables that shows the agreement/disagreement between Duke readers, values from Most dataset, and grades predicted by deep learning algorithm are located in appendix D.

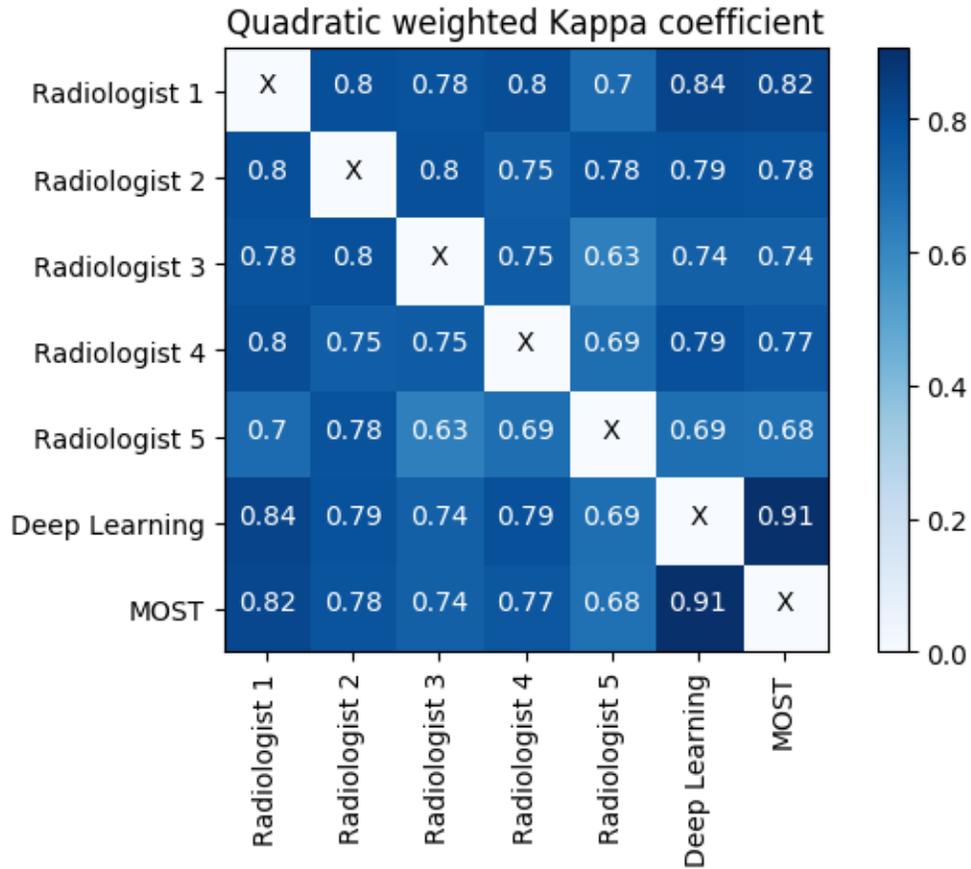

Figure 4. Quadratic weighted Kappa coefficient

# Discussion

In this study, we developed a deep learning model for fully automatic assessment of severity of knee osteoarthritis according to the Kellgren-Lawrence grading scheme. We compared our model to KL score assessment available in the MOST dataset as well to an assessment of 5 radiologists at our institution.

We conclude that the developed algorithm performs at the level of a radiologist since the difference of the algorithm's assessment from the radiologist's assessment was virtually the same as the differences between different radiologists. While this shows the strength of our algorithm, the only moderate agreement between radiologists underscores the limitations of the KL grade for measurement of severity of osteoarthritis. Our algorithm could provide a solution to this issue since it provides a score that is free of inter-reader variability (the same cases is always scored the same) and inter-reader variability (all cases are read by the same algorithm rather than multiple radiologists).

We observed the highest difference between our algorithm and the radiology readers for lower KL grades and particularly for grade 1. We hypothesize that this is caused by a higher level of subjectivity for these grades which is confirmed by a higher difference between different radiologists for those grades as well.

An assessment of an algorithm in a presence of a high inter-reader variability of the gold standard is challenging and is a frequent difficulty in radiology research studies. Once the algorithm reaches a certain level of performance, the difference between the algorithm's prediction and a radiologist's assessment can mean an error on the part of the algorithm, an error on the part of a radiologist, or that there is an inherent level of subjectivity present in the assessment. To address this issue to a certain extent, we conducted a reader's study to show that the difference between the algorithm and radiologists is similar to the differences between the radiologists themselves. This allowed us to draw a conclusion that the algorithm performs at the level of a radiologist given the level of uncertainly present in the task. Future studies could evaluate the effectiveness of the KL grade assessment, given by radiologists and the algorithm, in predicting outcomes of the knee replacement surgery.

Some prior efforts have taken place to tackle the task of knee OA assessment. The prior algorithms were based only on the PA views and did not offer comparison with multiple radiologists and therefore not allowing to place performance of the algorithm in a proper clinical context. Furthermore, our algorithm outperformed all prior algorithms according the evaluation presented in the prior studies. The table with a comparison between our results and results from prior publications are shown in table 5.

We observed a slight improvement in performance of the two-view model as compared with the one view model. The improvement, however, was small. This is consistent with the fact that vast majority of the information needed to assess KL score is present in the PA view and that KL score is often assessed using the PA view only. However, it would be difficult to conclude from this study that there is no information about osteoarthritis in the lateral view. First, the conclusions that we can draw apply specifically to the algorithm used in this study and it is a possibility that another algorithm would show a larger difference. Second, a proper comparison would include radiologists using that PA and lateral views together as well as the PA view alone. This data was not available in this study. All this said, this study provides some evidence in the direction that the information in the lateral views is limited.

In conclusion, the algorithm proposed in this study could provide an accurate and reproducible measure of osteoarthritis severity for research and clinical decision-making.

----

**Appendix A** Single input model architecture

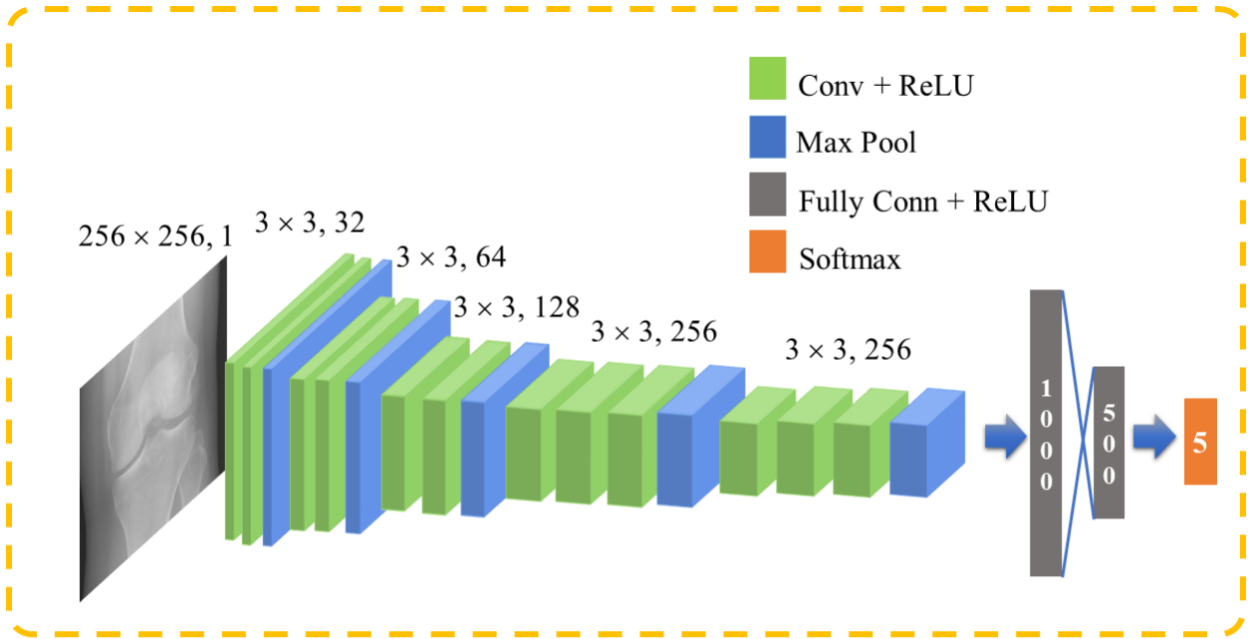

Figure X. Single input model architecture.

**Appendix B** Single view models' results

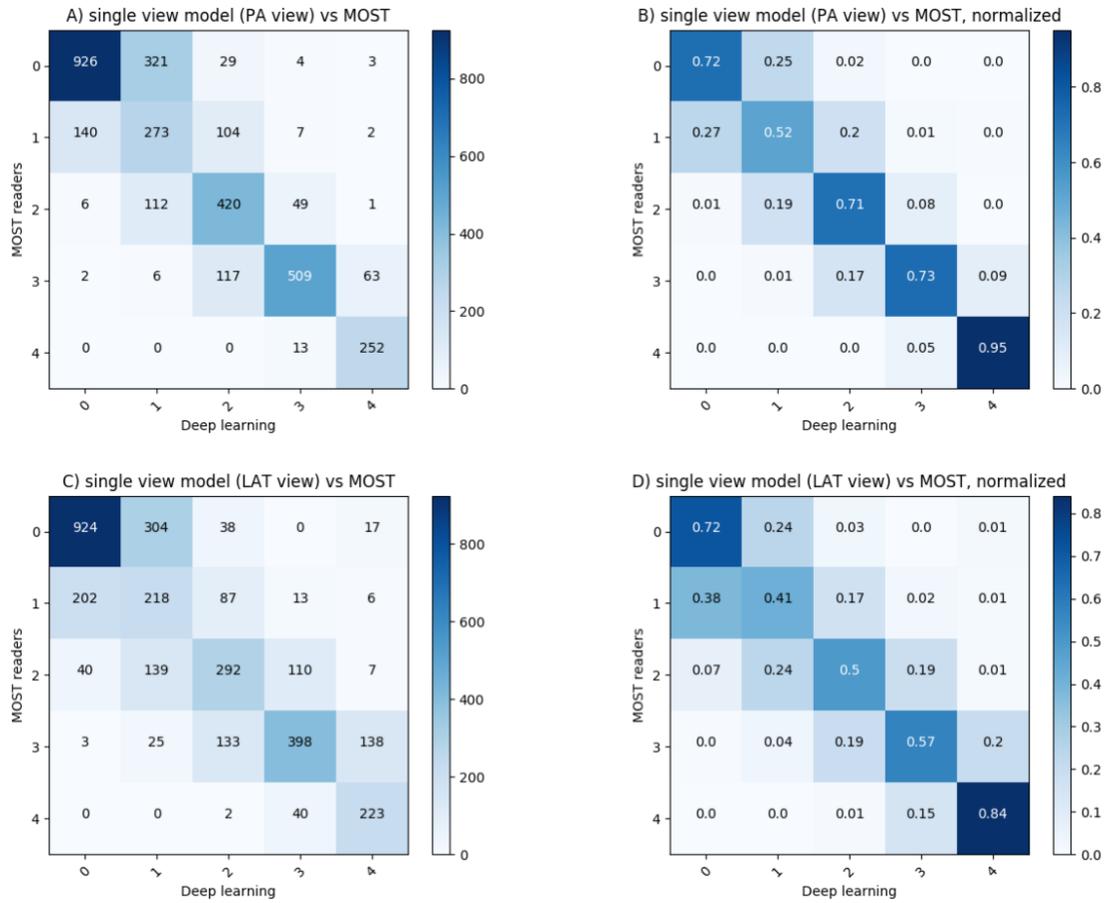

Figure Y. Confusion matrices for single input model

**Appendix C** non-weighted and linear weighted Kappa coefficient

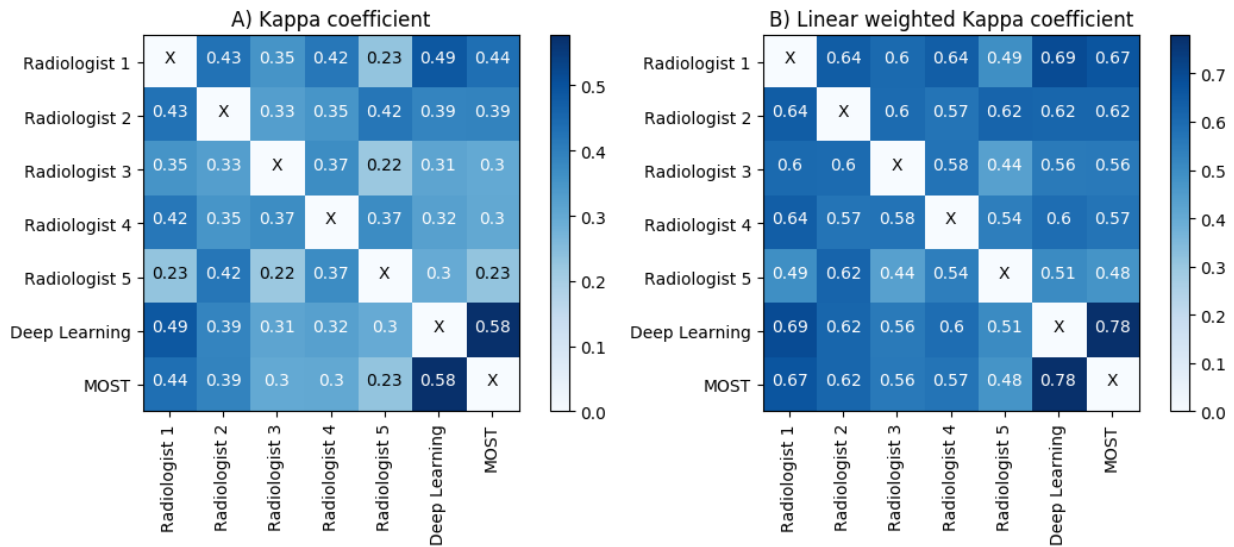

Figure Z. A) non-weighted kappa coefficient b) linear weighted Kappa coefficient

Appendix D Confusion matrixes between Duke readers, MOST assignments and deep learning.

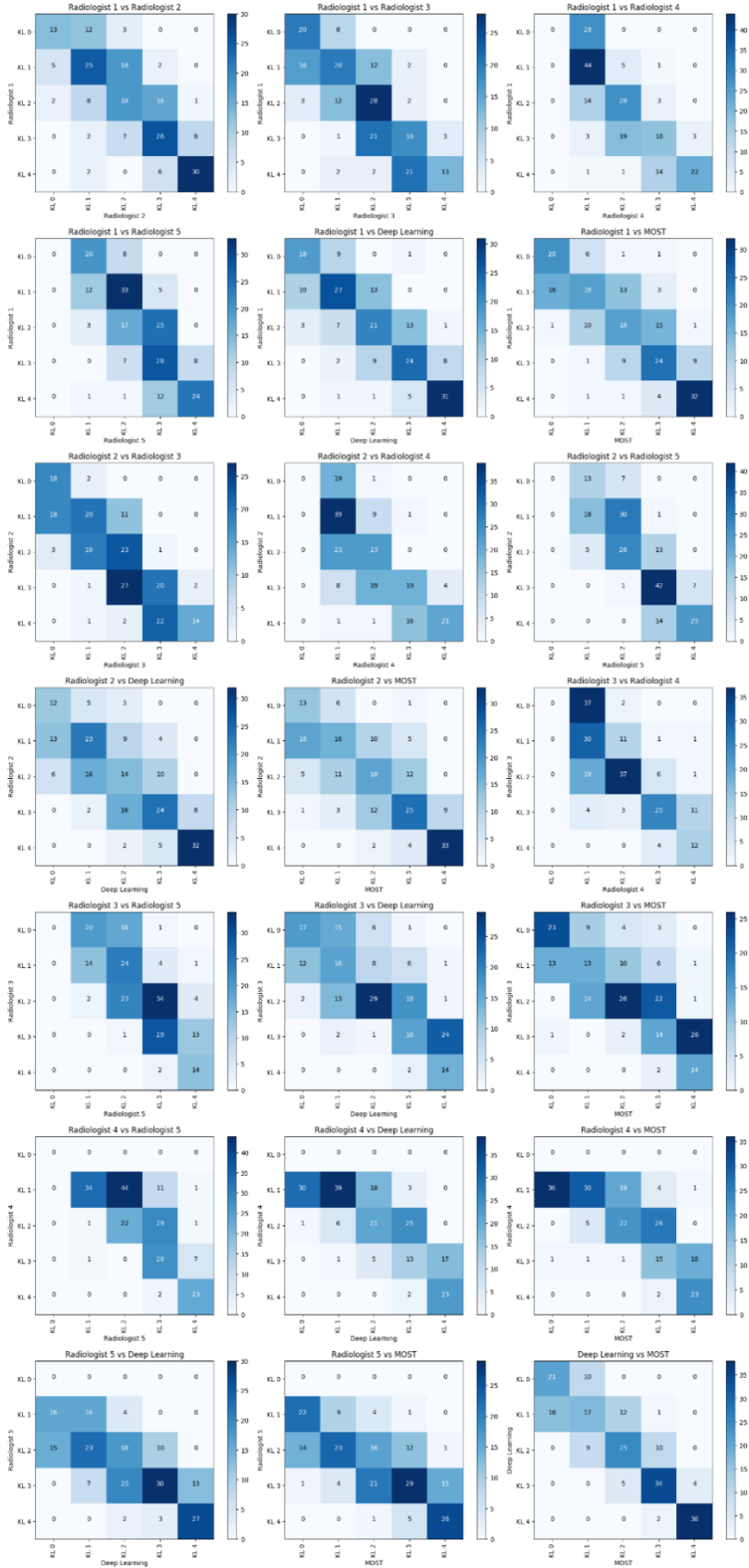

Figure Z2. Confusion matrixes between Duke readers, MOST assignments and deep learning.